\documentclass[aps,nofootinbib,prb,twocolumn,amsmath,amssymb,mathtools,showpacs,superscriptaddress]{revtex4-1}
\usepackage{graphics}
\usepackage{graphicx}
\usepackage{dcolumn}
\usepackage{bm}
\usepackage{color} 
\definecolor{darkblue}{rgb}{0,0,1.0}
\definecolor{darkred}{rgb}{0.6,0,0}
\definecolor{darkgreen}{rgb}{0,0.6,0}

\usepackage[colorlinks=true,urlcolor=darkblue,citecolor=darkblue,linkcolor=darkred,hyperfootnotes=false]{hyperref}
\makeatother

\begin{document}

\title{Defect removal by solvent vapor annealing in thin films of lamellar diblock copolymers}

\author{Xinpeng Xu}
\email{E-mail address: xu.xinpeng@gtiit.edu.cn}
\affiliation{Physics Program, Guangdong Technion -- Israel Institute of Technology, Shantou, Guangdong Province 515063, P. R. China}
\author{Xingkun Man}
\email{E-mail address: manxk@buaa.edu.cn}
\affiliation{Center of Soft Matter Physics and Its Applications, Beihang University, Beijing 100191, P. R. China}
\affiliation{School of Physics and Nuclear Energy Engineering, Beihang University, Beijing 100191, P. R. China}
\author{Masao Doi}
\affiliation{Center of Soft Matter Physics and Its Applications, Beihang University, Beijing 100191, P. R. China}
\affiliation{School of Physics and Nuclear Energy Engineering, Beihang University, Beijing 100191, P. R. China}
\author{Zhong-can Ou-Yang}
\affiliation{CAS Key Laboratory of Theoretical Physics, Institute of Theoretical Physics, Chinese Academy of Sciences, Beijing 100190, China}
\author{David Andelman}
\affiliation{Raymond and Beverly Sackler School of Physics and Astronomy, Tel Aviv University, Ramat Aviv 69978, Tel Aviv, Israel}

\begin{abstract}
  Solvent vapor annealing (SVA) is known to be a simple, low-cost and highly efficient technique to produce defect-free diblock copolymer (BCP) thin films. Not only can the solvent weaken the BCP segmental interactions, but it can vary the characteristic spacing of the BCP microstructures. We carry out systematic theoretical studies on the effect of adding solvent into lamellar BCP thin films on the defect removal close to the BCP order-disorder transition. We find that the increase of the lamellar spacing, as is induced by addition of solvent, facilitates more efficient removal of defects. The stability of a particular defect in a lamellar BCP thin film is given in terms of two key controllable parameters: the amount of BCP swelling and solvent evaporation rate. Our results highlight the SVA mechanism for obtaining defect-free BCP thin films, as is highly desired in nanolithography and other industrial applications.
\end{abstract}

\maketitle

\section{Introduction \label{sec:introduction}}

Thin films of diblock copolymer (BCP) with extended and defect-free lateral order can serve as ideal templates or scaffolds for fabricating nanoscale functional materials, as is highly desired in nanolithography and ultrafiltration membrane applications~\cite{Albert2010,Sinturel2013,Li2015}. However, when a BCP thin film coats a solid substrate, it is usually kinetically trapped in a nonequilibrium and poorly ordered state with significant amount of defects, hindering the mainstream use of BCP thin films in nanotechnology~\cite{Albert2010,Sinturel2013,Li2015}. In order to facilitate the fabrication of defect-free films, further treatments are needed. As examples to techniques that have been developed to tailor the self-assembly behavior of BCP thin films we mentioned electric field alignment~\cite{Russell1996,Russell2000}, shear alignment~\cite{Marencic2007,Shelton2017}, microwave annealing~\cite{Zhang2010}, and thermal annealing~\cite{Albert2010,Li2015,Li2014,Hur2015,Hur2018,Muller2016,Muller2018,Song2018,Man2012,Man2016}. Besides these approaches, solvent vapor annealing (SVA)~\cite{Albert2010,Sinturel2013,Baruth2014,Zhang2013,Zhang2014,Rudov2013,Gu2014a,
Gu2014b,Gu2016,KimLibera1998,Cavicchi2007,Fredrickson2013,Fredrickson2016,Albert2011,Hur2014,Hannon2015,Hao2017} has been used to enhance the mobility of polymer chains, and to facilitate the annihilation of defects.

In a typical SVA process~\cite{Albert2010,Sinturel2013,Hur2014,Zhang2014,Gu2014a,Gu2014b}, the BCP thin film is first exposed to vapors of one or more solvents for certain amount of annealing time. After a swollen and mobile polymer film is formed on the substrate, it is then dried by controlled solvent evaporation. During the film swelling in SVA, solvent molecules diffuse into the thin BCP film and screen (or weaken) the unfavorable interactions between polymer segments, leading to a lower effective Flory-Huggins parameter, $\chi_{\rm eff}$, than its value for dry (no solvent) case~\cite{Helfand1972,Lodge1995,Naughton2002,Gu2014a,Gu2014b}. Hence, the effective Flory-Huggins parameter $\chi_{\rm eff}=\chi_{\rm eff}(\phi)$ is a function of solvent volume fraction, $\phi$. Furthermore, as solvent is introduced, the spacing $\lambda=\lambda(\phi)$ of the microdomains also changes as a function of $\phi$~\cite{Shibayama1983a,Shibayama1983b,Sinturel2013,Hur2014,Baruth2014,
Zhang2013,Zhang2014,Rudov2013,Gu2014a,Gu2014b,Gu2016}.

Comparing to conventional thermal annealing (TA), SVA offers several unique advantages for defect annihilation~\cite{Albert2010,Sinturel2013,Gu2014a}: (i) SVA provides means to anneal BCPs that are sensitive to thermal degradation. (ii) SVA is generally more effective (with shorter annealing time) in removing defects for thin films of high-molecular-weight BCPs. (iii) SVA provides additional controllable parameters for tuning film morphology. For example, solvent molecules allow the segregation parameterized by $N\chi$ to vary continuously over a larger range than what can be achieved in TA. In addition, selective solvent can be used to induce asymmetric swelling between BCP blocks, and thus change the morphology of the BCP film.

Despite the widespread use of SVA, a quantitative understanding of its effect in defect removal has not yet been established~\cite{Albert2010,Sinturel2013,Gu2014a,Gu2014b,Hur2014,Baruth2014,Hannon2015,Hao2017,Fredrickson2013,Fredrickson2016}. The two principal problems are lack of standardized SVA processes and no rigorous understanding of the evolution of the nanostructure of the final dried film from
 its solvent-swollen state. Therefore, it is of interest to investigate theoretically how relevant system parameters affect the BCP swelling and solvent removal in SVA~\cite{Gu2014b,Hur2014,Hannon2015,Hao2017,Fredrickson2013}. Such parameters include swelling ratio (\emph{i.e.}, the fractional increase in the film volume due to solvent absorption), solvent-annealing time, solvent evaporation rate, film thickness, temperature, and solvent selectivity. The understanding of their effects on the structure of BCP thin films is necessary to fine-tune and control the SVA process in order to obtain the desired morphology.

\begin{figure}[htbp]
\centering
  \includegraphics[clip=true, viewport=1 1 750 450, keepaspectratio, width=0.42\textwidth]{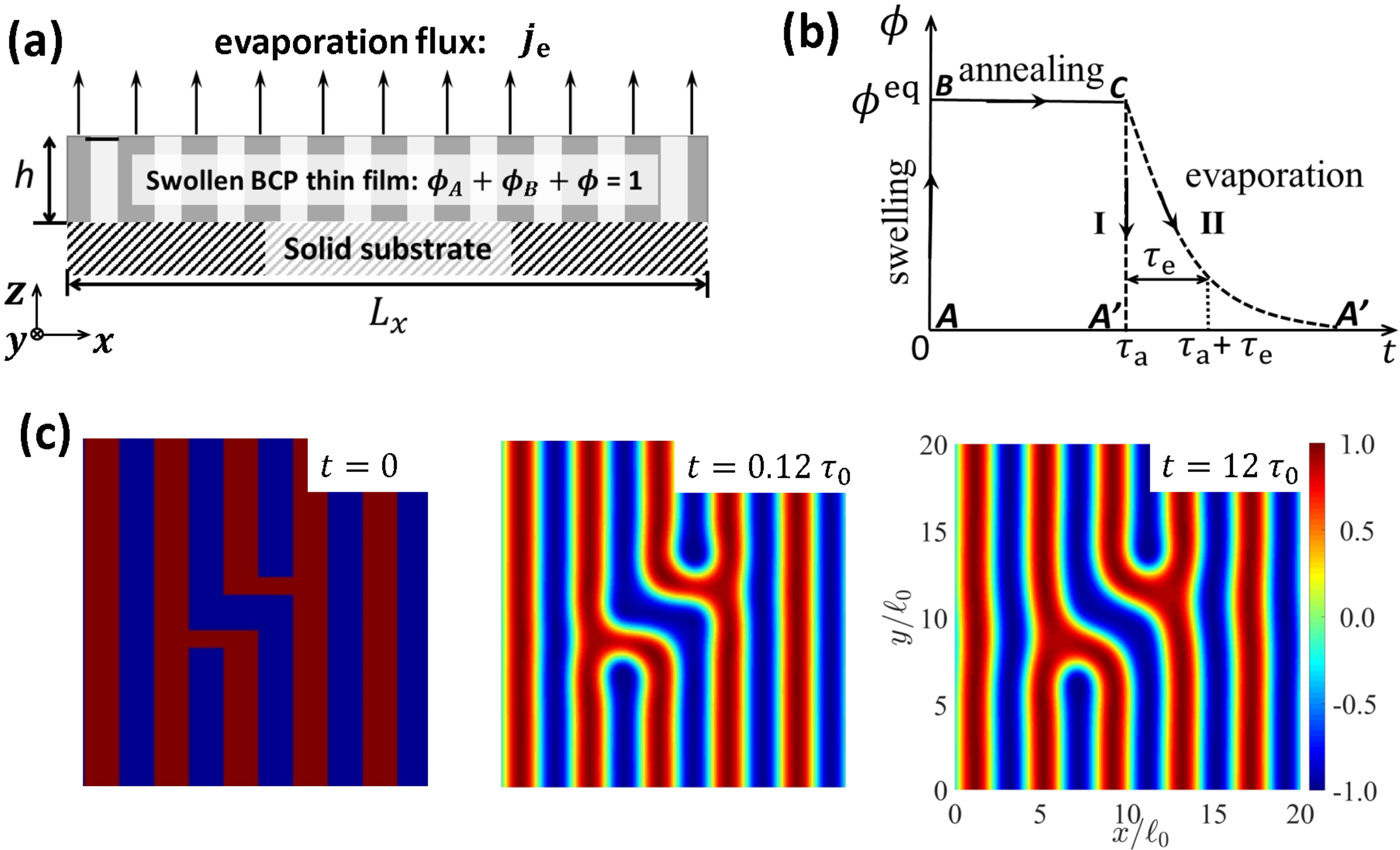}
  \caption {(a) Schematic illustrations of the BCP film setup. (b) A sketch of the temporal evolution of solvent volume fraction. Three important parameters are shown: the equilibrium solvent volume fraction, $\phi^{\rm eq}$, of a swollen film, the annealing time $\tau_{\rm a}$, and the characteristic evaporation time $\tau_{\rm e}$. The lines $A\rightarrow B$, $B\rightarrow C$ and $C\rightarrow A'$ represent the subsequent swelling, annealing and evaporation (deswelling) processes, respectively. (c) Preparation of the initial metastable defect structure from an artificially prescribed morphology at $t=0$, and its temporal evolution to $t=12\tau_0$.
  } \label{Fig:Schematic}
\end{figure}

Most previous theoretical and computational studies use either self-consistent field theory or particle-based simulation methods that include a large number of molecular parameters~\cite{Hur2014,Hannon2015,Fredrickson2013,Fredrickson2016,Hao2017}. Here, we propose an alternative continuum two-fluid model for SVA processes in BCP thin films. The detailed numerical analysis are carried out employing dimensionless system parameters: (i) Swelling ratio $\cal R$, representing the ratio of the swollen film volume $V$ to its original volume $V_0$, {\it i.e.,} ${\cal R}\equiv V/V_0$. (ii) Normalized solvent evaporation rate $\alpha_{\rm e} \equiv \tau_0/\tau_{\rm e}$ with $\tau_{\rm e}$ representing the characteristic time for complete solvent evaporation in the swollen film and $\tau_0$ characterizing the time for copolymer reorganization in the scale of microdomain period $\lambda$. More specifically, we investigate the removal of a typical defect occurring in symmetric BCP lamellae in their perpendicular orientation by SVA using non-selective solvent. In SVA experiments of thin BCP films~\cite{Gu2014a,Gu2014b,Baruth2014}, it has been well established that the swelling ratio, ${\cal R}$, and the solvent evaporation rate, $\alpha_{\rm e}$, are the essential annealing parameters and should be specified explicitly. However, a quantitative understanding of their effects on the directed self-assembly using SVA and, particularly, on the efficiency of SVA in defect removal is still missing. Our main theoretical finding is that the lamellar spacing increase induced by added solvent facilitates a more efficient removal of defects in BCP films. Moreover, the dependence of the final morphology of the dried BCP film on $\cal R$ and $\alpha_{\rm e}$ is summarized in a stability diagram, which can be verified in future experiments.

In the next section, we present the thermodynamical modeling of BCP solutions and two-fluid model for evaporating BCP solution films. A simplified two-dimensional model is obtained in the thin-film limit without macroscopic convection. Our numerical method and the preparation of a specific initial defect are also explained. In section III, we show and discuss our simulation results. Focus is placed on the effects of solvent-swelling ratio and solvent-evaporation rate on the removal of the initial defect in the BCP film. This paper is then summarized and concluded in section IV. Some qualitative remarks are given in the end.

\section{Model and Methods \label{sec:model}}

\subsection{Thermodynamics of BCP solutions: Generalized Landau-Brazovskii theory \label{sec:energy}}

When a BCP melt is exposed to the vapor of a good solvent, solvent molecules diffuse into the melt to form a swollen BCP solution, in which the chains gain mobility. We consider a BCP solution of monodisperse A/B diblocks dissolved in a {\it neutral} (non-selective) good solvent for which the two polymer-solvent interactions are equal, $\chi_{\rm AS}=\chi_{\rm BS}$. We restrict ourselves to symmetric A/B BCP where an A-block with $N_{\rm A}$ monomers is covalently bonded to a B-block with $N_{\rm B}$ monomers such that $N_{\rm A}=N_{\rm B}=N/2$, and $N$ being the total chain length (polymerization index). For simplicity, we assume incompressibility upon mixing the solvent and the BCP. In this case, the structure of the BCP solution can be described by two independent order parameters: the solvent volume fraction $\phi$ with $0\leq\phi\leq1$ and the relative volume fraction of the two blocks
\begin{align} \label{eq:order}
\psi \equiv \phi_{\rm A}-\phi_{\rm B},
\end{align}
with $-1\leq\psi \leq1$. In addition, since the BCP solution is incompressible, we have
\begin{align} \label{eq:incomp}
\phi_{\rm A}+\phi_{\rm B}+\phi=1, \quad {\rm or}  \quad \phi_{\rm P}+\phi=1,
\end{align}
in which $\phi_{\rm P} = \phi_{\rm A}+\phi_{\rm B}$ is the total volume fraction of A/B copolymer blocks. Another simplifying assumption in the above is that the two monomers have the same molecular volumes.

It has been known that adding a neutral solvent to the BCP melt has a similar effect as increasing the temperature~\cite{Gu2014a,Gu2016}. Namely, shifting the phase diagram of BCP melts towards higher temperature~\cite{Lodge1998}. Based on this understanding, a symmetric BCP solution close to the order-disorder transition (ODT) can be described by generalizing the Landau-Brazovskii free-energy ${\cal F}$ ~\cite{Binder1994,Majewski2016,David1997,David2002,Ciach2013,Brazovskii1975,OhtaIto1995} (in units of $k_{\rm B}T$) as
\begin{align} \label{eq:Fb}
{\cal F}[\psi,\phi]&= \int {\rm d}^3 \mathbf{r} \left[f_{\rm S}(\phi)+\frac{ \tau}{2} \psi^2 +\frac{u}{4!} \psi^4+ \frac{\kappa}{2}\left(\nabla^2 \psi + q^2 \psi\right)^2 \right]\;,
\end{align}
in which the ideal mixing entropy of the solvent is $f_{\rm S}(\phi)=\phi\ln \phi-\phi$, and the parameters are given by~\cite{Binder1994,Majewski2016,David1997,David2002,Ciach2013,Brazovskii1975}
\begin{align}\label{eq:Fbsuba}
\tau \simeq 2n_c  N (\chi_{\rm c}-\chi_{\rm eff}) , \; u \simeq n_c, \; \kappa\simeq \frac{6 n_c}{q^{4}},\; q \sim  \frac{1}{R_{\rm g}},
\end{align}
with $n_c =1/(Na^3)$ being the copolymer closed-packing chain density of an incompressible BCP melt, $a$ the monomer size, $N\chi_{\rm c} \simeq 10.5$ for the melt ODT (no solvent), $R_{\rm g}^2 = Na^2/6$ the gyration radius for Gaussian chains, and $q$ is the characteristic wavenumber of ordered BCP microdomains, which gives the thermodynamically-preferred lamellar spacing $\lambda=2\pi/q \sim R_{\rm g}$.

For a swollen BCP solution with nonselective solvent, the effective interaction parameter $\chi_{\rm eff}$ in $\tau$ of eq.~(\ref{eq:Fbsuba}) and the wavenumber $q$ (or the lamellar spacing $\lambda$) are both functions of the solvent volume fraction, $\phi$. That is, the phase diagram of BCP solutions is described by the same form of energy functional~\eqref{eq:Fb} as BCP melts, in which the effects of added solvent are represented by solvent-dependent $\chi_{\rm eff}(\phi)$ and $q(\phi)$.

For $\chi_{\rm eff}$, we follow the experimental results and take the following power-law form~\cite{Helfand1972,Lodge1995,Naughton2002,Gu2014a,Gu2014b},
\begin{equation} \label{eq:chieff}
\chi_{\rm eff}= \chi (\phi_{\rm P}) ^{\delta}= \chi (1-\phi) ^{\delta},
\end{equation}
in which the exponent $\delta>0$ varies from $1.0$ to $1.6$ in a wide range of $\phi$~\cite{Helfand1972,Lodge1995,Naughton2002}. This $\chi_{\rm eff}(\phi_S)$  represents the dilution effect of added solvent on the polymer chains by reducing the unfavorable A/B segmental interactions. Particularly, the case with $\delta=1.0$ can be obtained from the mean-field \emph{dilution approximation}\cite{Helfand1972,Lodge1995}, by assuming a uniform distribution of solvent throughout the BCP structure.
The scaling of $\lambda$ with the volume fraction of solvent is often approximated by another power law~\cite{Shibayama1983a,Shibayama1983b,Sinturel2013,Hur2014,Baruth2014,
Zhang2013,Zhang2014,Rudov2013,Gu2014a,Gu2014b,Gu2016}
\begin{equation}\label{eq:spacing1}
\lambda= \lambda_0 (\phi_{\rm P})^{\beta}=\lambda_0 (1-\phi)^{\beta},
\end{equation}
which represents the effect of added solvent on the BCP lamellar spacing $\lambda$ with $\lambda_0$ being the lamellar spacing of the dry film. Here, the exponent
$\beta\lessgtr 0$ is a characteristic parameter of BCP solutions~\cite{Shibayama1983a,Shibayama1983b,Sinturel2013,Hur2014,Baruth2014,Zhang2013,Zhang2014,Rudov2013,Gu2014a,Gu2014b,Gu2016}, and depends in general on the degree of solvent selectivity, solvent volume fraction, and morphology of the ordered state. Hence, the wavenumber $q$ in the free energy (\ref{eq:Fb}) is given by
\begin{equation}\label{eq:q}
q=q_0 (1-\phi)^{-\beta},
\end{equation}
with $q_0\equiv q(\phi=0)=2\pi/\lambda_0$.

It is known from previous studies both in BCP bulk solutions ~\cite{Shibayama1983a,Shibayama1983b} and films~\cite{Sinturel2013,Gu2014a,Gu2014b,Gu2016} that the solvent-induced changes in lamellar spacing can be classified into two regimes according to the solvent volume fraction, $\phi$, with a crossover between them at $\phi=\phi^{*}$, where $\phi^{*}$ is taken in our theory to be an external (material) parameter.

\emph{Kinetically controlled regime} with $\beta<0$ for low amount of solvent, $\phi<\phi^{*}$. Here, the lamellar spacing $\lambda$ increases with increasing $\phi$, and $\lambda$ is kinetically controlled~\cite{Shibayama1983a,Shibayama1983b,Gu2016} in the sense that the copolymers do not gain much motility. The BCP structure reorganization cannot take place in a typical annealing time, and further addition of solvent therefore results in an expansion in all directions. For example, it can be obtained from solvent volume conservation that for an isotropic swelling $\beta = -1/3$, and for a uniaxial swelling of a (dense) BCP solution $\beta = -1$.

\emph{Thermodynamically controlled regime} with  $\beta>0$ for larger amount of solvent, $\phi>\phi^{*}$. Here, the lamellar spacing decreases with increasing $\phi$. This results from a decrease in segregation magnitude as seen in eq.~(\ref{eq:chieff})~\cite{Sinturel2013,Gu2014a,Gu2014b,Gu2016}. For large $\phi$, $\lambda$ is thermodynamically controlled~\cite{Shibayama1983a,Shibayama1983b,Gu2014a,Gu2014b,Gu2016} in the sense that the copolymers are sufficiently mobile and expand efficiently along the lamellae interface for a given annealing time. Any addition of solvent can then result in a decrease of the lamellar spacing $\lambda$, in a direction \emph{perpendicular} to the interface.

For example, it was found experimentally that $\beta \approx 1/3$ for dilute BCP bulk solutions of PS-PI with nonselective solvents~\cite{Shibayama1983a,Shibayama1983b}, and $\beta \approx 2/3$ for dilute solution films of polystyrene-poly(2-vinyl pyridine) (PS-P2VP) with nonselective solvents~\cite{,Gu2014a,Gu2014b}.
The crossover solvent volume fraction $\phi^{*}$, is found experimentally to be, $\phi^{*}\approx 0.3$ for polystyrene-polyisoprene (PS-PI) and PS-P2VP copolymers in solutions with non-selective solvent~\cite{Shibayama1983a,Gu2014a,Gu2014b}.

The equilibrium structure of a swollen BCP solution is then given by the minimization of the Landau-Brazovskii free-energy $\mathcal F$ of eq.~(\ref{eq:Fb}) with respect to $\psi$ and $\phi$, while subject to the constraints of conserved $\phi$ and phase parameter $\psi$:
\begin{eqnarray}
\label{eq:equil-muhat1}
\mu_{\rm S} &\equiv &\frac{\delta {\mathcal F}}{\delta \phi} =\ln \phi= {\mathrm{const.}}, \\
\label{eq:equil-muhat2}
\mu_{\psi} &\equiv& \frac{\delta {\mathcal F}}{\delta \psi} = \tau(\phi) \psi + u\psi^3 + \kappa \left[q^4(\phi) \psi \right. \nonumber \\
&&\quad \quad\quad\left. +2 q^2(\phi) \nabla^2 \psi + \nabla^4 \psi\right] = {\mathrm{const.}}
\end{eqnarray}
Note that in eq.~(\ref{eq:equil-muhat1}) we have neglected the $\phi-$dependent contribution of $\tau$ and $q$ as shown in eqs.~\eqref{eq:Fbsuba}, ~\eqref{eq:chieff}, and \eqref{eq:q}. This is equivalent to neglecting the influence of copolymers on the chemical equilibrium of solvent, $\mu_{\rm S}$, on the other hand, $\phi$ appears in eq.~\eqref{eq:equil-muhat2} as $\tau$ and $q$ are both functions of $\phi$. The resulting equilibrium distribution is consistent with the assumption of uniform distribution of solvent through the BCP films. In addition, we would like to point out that the Landau-Brazovskii free-energy~(\ref{eq:Fb}) can be further extended to include more physical effects by introducing more terms, \emph{e.g.}, solvent selectivity via a coupling term $\phi\psi$, solvent accumulation at A/B interface via a gradient term $(\nabla \phi)^2$, and Marangoni effects via a concentration-dependent rigidity $\kappa(\phi)$.

Finally, the bulk eqs.~\eqref{eq:equil-muhat1} and ~\eqref{eq:equil-muhat2} is supplemented with boundary conditions given at the neutral (non-selective) solid surface by
\begin{eqnarray}
\hat{\mathbf n} \cdot \nabla \psi &=&0,\nonumber\\
\hat{\mathbf n} \cdot \nabla (\nabla^2 \psi + q^2\psi)&=&0,
\label{eq:equil-bc1}
\end{eqnarray}
and at the film-air interface
\begin{eqnarray}
\nabla^2 \psi + q^2 \psi &=&0, \nonumber \\
\hat{\mathbf n} \cdot \nabla (\nabla^2 \psi + q^2\psi)&=&0.
\label{eq:equil-bc3}
\end{eqnarray}
Here $\hat{\mathbf n} = (-\nabla_{\parallel} h, 1)/{\sqrt{1+(\nabla_{\parallel} h)^2}}$ is the normal unit vector of the film-air interface with $h$ being the film thickness and $ \nabla_{\parallel}$ being the 2D gradient operator in the $xy$-plane.

Close to the ODT (where the order-disorder transition occurs), the BCP solution is described by two lengths: the periodic spacing of BCP lamellae $\lambda=2\pi/q$, and the correlation length $\xi\sim (\tau/\kappa)^{-1/4}\sim \lambda (N\chi_{\rm eff} -N\chi_{\rm c})^{-1/4}$, which diverges at $\chi_{\rm eff} =\chi_{\rm c}$ (the ODT). In particular, for BCP solution films below the ODT ($\tau<0$ or $N\chi_{\rm eff} >N\chi_{\rm c}$), a defect-free perpendicular lamellar state can be described in the single-mode approximation (in the limit of weak segregation, $N\chi_{\rm eff}  \geq N\chi_{\rm c}$) by
\begin{equation}\label{eq:perfect}
\psi=\psi_{q} \cos (qx),
\end{equation}
with $\psi_{q}\sim \sqrt{8\tau/u}=4(N\chi_{\rm eff} - N\chi_{\rm c})^{1/2}$ being the amplitude obtained by minimizing the free energy with respect to $\psi_q$, and $\lambda=2\pi/q$ is the periodic spacing of the lamellae.


\subsection{Dynamics of BCP solutions: Two-fluid model \label{sec:model-twofluid}}

With the above described generalized Landau-Brazovskii free-energy, we formulate a two-fluid model for the BCP solutions with non-selective solvent to explore the dynamic coupling between BCP copolymers and small solvent molecules during SVA. This is in analogy to a two-fluid model used to model polymer solutions, polymer blends, and diblock copolymer melts~\cite{Onuki1990, DoiOnuki1992,Milner1993,Buxton2007,Vinals2012,Yabunaka2012}.

Let us consider a BCP solution in which bulk flow and diffusion are taking place simultaneously, \emph{i.e.}, the velocity of the A/B blocks differ from each other as well as from the solvent velocity. From the conservation of solvent and A/B blocks we have
\begin{align} \label{eq:phij}
\partial_t \phi_{j} = -\nabla \cdot (\phi_{j}{\mathbf v}_{j}),
\end{align}
where $j={\rm A},\, {\rm B}\,$ or ${\rm S}$ represents the three components: A-block, B-block, or solvent, respectively, and ${\mathbf v}_{j}$ and $\phi_{j}$ the velocity and volume fraction of each component, respectively. For simplicity, we assume incompressibility as given in eq.~\eqref{eq:incomp} and zero volume change upon mixing solvent in the BCP solution, from which we obtain
\begin{align} \label{eq:divv}
\nabla \cdot \bar{\mathbf v}=0.
\end{align}
with the volume averaged velocity $\bar{\mathbf v}$ defined by $\bar{\mathbf v}\equiv \phi{\mathbf v}_{\rm S}+\phi_{\rm A} {\mathbf v}_{\rm A}+\phi_{\rm B} {\mathbf v}_{\rm B}$.

From eqs. (\ref{eq:phij}) and (\ref{eq:divv}) we obtain
\begin{align} \label{eq:phisphi} 
{\partial_t  \phi} + \bar{\mathbf v}\cdot \nabla\phi&=-\nabla\cdot {\mathbf j}_{\rm S}, \nonumber \\
{\partial_t  \psi}+ \bar{\mathbf v}\cdot \nabla\psi &= -\nabla \cdot {\mathbf j}_{\psi},
\end{align}
where
\begin{align} \label{eq:phiAB} 
\phi_{\rm A}&=(1-\phi+\psi)/2, \nonumber \\
\phi_{\rm B}&=(1-\phi-\psi)/2, \nonumber \\
{\mathbf j}_{\rm S}&\equiv \phi \left({\mathbf v}_{\rm S}-\bar{\mathbf v}\right) = -\phi_{\rm A} \left({\mathbf v}_{\rm A}-\bar{\mathbf v}\right)-\phi_{\rm B} \left({\mathbf v}_{\rm B}-\bar{\mathbf v}\right), \nonumber \\
{\mathbf j}_{\psi} &\equiv \phi_{\rm A} \left({\mathbf v}_{\rm A}-\bar{\mathbf v}\right)-\phi_{\rm B} \left({\mathbf v}_{\rm B}-\bar{\mathbf v}\right),
\end{align}
where ${\mathbf j}_{\rm S}$ and ${\mathbf j}_{\psi}$ are two currents.
Note that $\phi$ and $\psi$ are both conserved order parameters of the BCP solution.

In order to determine $\bar{\mathbf v}$, ${\mathbf j}_{\rm S}$, and ${\mathbf j}_{\psi}$, we employ Onsager's variational principle~\cite{Onuki1990, DoiOnuki1992, Milner1993, Buxton2007, Vinals2012, Xu2015}, in which the Rayleighian functional is ${\mathcal R} = \dot {\mathcal F}+ {\Phi}$ with the temporal change rate of free energy given by
\begin{align} \label{eq:Fdot}
\partial_t{\cal F}= \dot {\cal F}= \int {\rm d}^3 {\mathbf r}(\mu_{\psi} {\partial_t \psi}+{\mu_{\rm S}} {\partial_t \phi}).
\end{align}
The dissipation function is generally given by
\begin{align} \label{eq:Phi0}
{\Phi}= &\int {\rm d}^3 {\mathbf r}\left[\sum_{i,j}\frac{\tilde{\zeta}_{ij}}{2} {\mathbf v}_i\cdot {\mathbf v}_j+ \frac{\zeta_{\rm S}}{2}({\mathbf v}_{\rm B}-{\mathbf v}_{\rm S})^2
+\frac{\eta}{4}  (\nabla \bar{\mathbf v}+\nabla \bar{\mathbf v}^T)^2\right],
\end{align}
where $i\,, j={\rm A, \, B,\, S}$ and the friction matrix $\tilde{\zeta}_{ij}$ must be symmetric. Furthermore, since ${\Phi}$ has to be invariant under Galilean transformation, we get
\begin{align} \label{eq:zeta}
\sum_{i}\tilde{\zeta}_{ij}=0, \quad
\sum_{j}\tilde{\zeta}_{ij}=0,
\end{align}
and hence there are only \emph{three} independent elements in the $\tilde{\zeta}_{ij}$ matrix. We can now rewrite the Galilean-invariant dissipation function in the form
\begin{widetext}
\begin{equation}\label{eq:Phi1}
\Phi= \int {\rm d}^3 {\mathbf r}\left[\frac{\zeta}{2} ({\mathbf v}_{\rm A}-{\mathbf v}_{\rm B})^2+ \frac{\zeta_{\rm S}}{2}({\mathbf v}_{\rm A}-{\mathbf v}_{\rm S})^2+ \frac{\zeta_{\rm S}}{2}({\mathbf v}_{\rm B}-{\mathbf v}_{\rm S})^2
+\frac{\eta}{4}  (\nabla \bar{\mathbf v}+\nabla \bar{\mathbf v}^T)^2\right],
\end{equation}
\end{widetext}
or equivalently
\begin{widetext}
\begin{equation} \label{eq:Phi2}
\Phi= \int {\rm d}^3 {\mathbf r}\left[\frac{1}{2} L_{\psi\psi}\, {\mathbf j}_{\psi} ^2+ \frac{1}{2} L_{\psi {\rm S}}\, {\mathbf j}_{\psi}\cdot {\mathbf j}_{\rm S} +\frac{1}{2} L_{{\rm S}\psi}\, {\mathbf j}_{\psi}\cdot {\mathbf j}_{\rm S}+\frac{1}{2} L_{\rm SS}\, {\mathbf j}_{\rm S} ^2 +\frac{\eta}{4}  (\nabla \bar{\mathbf v}+\nabla \bar{\mathbf v}^T)^2\right],
\end{equation}
\end{widetext}
where the resistance matrix $L_{ij}$ (with $i,\,j ={\rm S},\,{\rm \psi}$) is given by
\begin{equation}\label{eq:MatrixL}
{\mathbf L}\sim \left[
\begin{matrix}
L_{\rm SS} & L_{{\rm S}\psi} \\
L_{\psi {\rm S}} &L_{\psi\psi}
\end{matrix}
\right],
\end{equation}
with
\begin{widetext}
\begin{align}\label{eq:MatrixL}
L_{\psi\psi}&=
4\zeta (1-\phi)^2 + \zeta_{\rm S}\left[ (1-\phi+\psi)^2  + (1-\phi-\psi)^2\right], \nonumber \\
L_{\psi {\rm S}}= L_{{\rm S}\psi}&=
4\zeta \psi (1-\phi) + \frac{\zeta_{\rm S}}{\phi}\left[ (1-\psi) (1-\phi+\psi)^2 - (1+\psi) (1-\phi-\psi)^2 \right], \nonumber \\
L_{\rm SS}&=
4\zeta \psi^2 + \frac{\zeta_{\rm S}}{\phi^2}\left[ (1-\psi)^2(1-\phi+\psi)^2 +(1+\psi)^2(1-\phi-\psi)^2\right].
\end{align}
\end{widetext}
In the above, we have assumed equilibrium conditions as given in eqs.~\eqref{eq:equil-muhat1} and ~\eqref{eq:equil-muhat2} at the boundaries of the BCP film. The first three terms in the energy dissipation~(\ref{eq:Phi1})-(\ref{eq:Phi2}) are caused by the relative motion either between the two polymer blocks or between the polymer blocks and solvent. The fourth term represents the energy dissipation caused by solvent velocity gradients, where $\eta$ is the solution viscosity, and $\zeta$ and $\zeta_{\rm S}$ are the friction constants per unit volume.
Note that for neutral solvents (\emph{i.e.,} non-selective solvents that interacts equally with A/B blocks), the same friction constant, $\zeta_{\rm S}$, is used to describe the relative motion between the solvent and the A/B blocks, respectively.
In addition, we would like to point out that the viscoelastic properties of BCP solutions can be incorporated into the two-fluid model as discussed in Ref.~[46] for polymer solutions and blends.

Substituting eq.~(\ref{eq:phisphi}) into eqs.~(\ref{eq:Fdot}) and (\ref{eq:Phi2}), minimizing Rayleighian with respect to $\bar{\mathbf v}$, ${\mathbf j}_{\rm S}$, and ${\mathbf j}_{\psi}$ and substituting the obtained results back into eq.~(\ref{eq:phisphi}), we obtain the following set of dynamic equations~\cite{Onuki1990, DoiOnuki1992, Milner1993, Buxton2007, Vinals2012, Xu2015}

\begin{align} \label{eq:dyn}
&{\partial_t  \phi} + \bar{\mathbf v}\cdot \nabla\phi=\nabla\cdot \left[M_{\rm SS} \nabla \mu_{\rm S} + M_{\rm S\psi} \nabla\mu_{\psi}\right], \nonumber \\
&{\partial_t  \psi}+ \bar{\mathbf v}\cdot \nabla\psi = \nabla \cdot\left[M_{\psi {\rm S}} \nabla \mu_{\rm S} + M_{\psi\psi} \nabla\mu_{\psi}\right],
\nonumber \\
&-\nabla p + \eta \nabla^2 \bar{\mathbf v}-\psi\nabla \mu_{\psi}-\phi\nabla \mu_{\rm S}=0,
\end{align}
with the incompressibility condition (\ref{eq:divv}). Here, we have substituted the following constitutive relations resulted from the Rayleighian minimization
\begin{align} \label{eq:jSjphi}
&{\mathbf j}_{\rm S} =-\left[M_{\rm SS} \nabla \mu_{\rm S} + M_{\rm S\psi} \nabla\mu_{\psi}\right], \nonumber \\
&{\mathbf j}_{\psi}=-\left[M_{\psi {\rm S}} \nabla \mu_{\rm S} + M_{\psi\psi} \nabla\mu_{\psi}\right]
\end{align}
into the conservation eq.~\eqref{eq:phisphi}. The motility coefficient matrix $M_{ij}$ (with $i,\,j ={\rm S},\,{\rm \psi}$) is the inverse of the resistance matrix $L_{ij}$ given in eq.~\eqref{eq:MatrixL}. It is also symmetric and its elements in the limit of dilute solution (\emph{i.e.}, small $\phi$) are found to be
\begin{equation}\label{eq:Matrix}
{\mathbf M}\sim \left[
\begin{matrix}
\zeta_{\rm S}^{-1} \phi^2 &-\zeta_{\rm eff}^{-1} \phi \psi \\
-\zeta_{\rm eff}^{-1} \phi \psi &\zeta_{\rm eff}^{-1}
\end{matrix}
\right]
\end{equation}
where the effective friction constant per volume $\zeta_{\rm eff}$ depends on the relative values of $\zeta$ and $\zeta_{\rm S}$.

Note that here, for simplicity, we only consider the case of isotropic diffusion as an application of the two-fluid model. This means that all the mobility coefficients $M_{ij}$ are scalars. However, it is worthwhile to point out that the copolymer diffusion is in general anisotropic, depending on the BCP morphology~\cite{Hur2014,Vinals2012,Yokoyama2006}. Namely, each motility coefficient $M_{ij}$ should be an anisotropic tensorial function of the local morphology. It has been shown that anisotropic diffusion can significantly affect the rate of relaxation of perturbations in BCP films~\cite{Vinals2012,Yokoyama2006}. Particularly for lamellar BCP films, the effects of transverse mobility are generally negligible as compared to longitudinal copolymer diffusion. Thus, we expect that when a defect becomes unstable, such anisotropic diffusion in a lamellar BCP thin film can significantly increase the longitudinal boundary velocities of defects through the coupling between the longitudinal and transverse diffusive modes within the defect region.

For a thin film of BCP solution, the above set of equations has to be supplemented with proper boundary conditions~\cite{Xu2015}. At the bottom (neutral) substrate ($z=0$), we take the no-slip condition for the average velocity and the impermeability conditions for diffusion fluxes as
\begin{align}\label{eq:bc0}
&\bar{\mathbf v}=0, \nonumber \\
&\hat{\mathbf n} \cdot{\mathbf j}_{\rm S} =\hat{\mathbf n} \cdot{\mathbf j}_{\psi}=0,
\end{align}
with ${\mathbf j}_{\rm S}$ and ${\mathbf j}_{\psi}$ given by eq.~\eqref{eq:jSjphi}, and assume equilibrium conditions for order parameter, $\psi$, as given in eq.~\eqref{eq:equil-bc1}.
At top free surface at $z=h(t)$, we have stress-free conditions
\begin{align}\label{eq:bch}
& \hat{\mathbf n} \cdot{\boldsymbol \sigma}=0,
\end{align}
where ${\boldsymbol \sigma}=-p {\mathbf I}+ \eta(\nabla \bar{\mathbf v}+\nabla \bar{\mathbf v}^T)$ is the viscous stress tensor.  We also assume equilibrium conditions for order parameter, $\psi$, as given in eq.~\eqref{eq:equil-bc3}.
In addition, at the free boundary $z=h(t)$, solvent evaporation occurs, and from the evaporation flux defined by $j_{\rm e}\equiv \rho_0(\bar{\mathbf v}-{\mathbf v}_{\rm i})\cdot \hat{n}=\rho_{\rm S}({\mathbf v}_{\rm S}-{\mathbf v}_{\rm i})\cdot \hat{n}$ we can get the kinematic boundary conditions as
\begin{eqnarray}
&\hat{\mathbf n} \cdot \bar{\mathbf v} = \partial_th +j_{\rm e}/\rho_0,\nonumber\\
&\hat{\mathbf n} \cdot {\mathbf j}_{\rm S} = (1-\phi)j_{\rm e}/\rho_0,\nonumber \\
&\hat{\mathbf n} \cdot {\mathbf j}_{\psi} = -\psi j_{\rm e}/\rho_0.\label{eq:bchk2}
\end{eqnarray}
Here the velocity of the free film/air interface satisfies ${\mathbf v}_{\rm i} \cdot \hat{n}=\partial_t h$. We assume that the evaporation flux follows the Hertz-Knudsen relation~\cite{Fredrickson2013,Fredrickson2016}
\begin{align}\label{bcje}
j_{\rm e}=\rho_0v_{\rm e}\left[\phi(h)-\phi^{\rm eq}\right],
\end{align}
in which $\phi^{\rm eq}$ is the equilibrium solvent fraction of a swollen film in coexistence with the vapor phase, $v_{\rm e}$ is an adjustable velocity parameter, proportional to mass transfer coefficient. It relates the rate of evaporative loss $\tau_{\rm e}^{-1}$ from the free surface to the solvent inter-diffusion through the film by $v_{\rm e}=h_0/\tau_{\rm e}$ with $h_0$ being the initial thickness of the BCP film.  As the surface tension is usually quite small in comparison to the bulk energy of the BCP film, we neglected its contribution. This results in a flat free surface of the BCP film throughout the SVA process.

Regardless of the physical properties of BCP films, the formulated problem is intrinsically non-linear due to the moving free boundary. Numerical integrations are carried out following a scheme that was successfully employed to other moving boundary problems~\cite{Xu2015,Tu1990,Crank1987}. The method consists of the following four steps:
(i)~Assume that $\psi({\mathbf r}, t)$, $\phi({\mathbf r}, t)$ and $\bar{\mathbf v}({\mathbf r}, t)$ are known functions at time $t$. The functions $\psi({\mathbf r}, t)$, $\phi({\mathbf r}, t)$ and $\bar{\mathbf v}({\mathbf r}, t)$ satisfy eqs.~\eqref{eq:dyn} and \eqref{bcje};
(ii)~Neglect the dynamic nature of $\psi({\mathbf r}, t)$, $\phi({\mathbf r}, t)$ and $\bar{\mathbf v}({\mathbf r}, t)$. The moving boundary $h(x,y,t)$ is extrapolated to the forward time step, $t + \Delta t$, by means of an explicit forward difference expression of eq.~(\ref{eq:bchk2});
(iii)~Let the boundary $h(x,y,t)$ be fixed at the new time step $t+\Delta t$. Determine $\psi({\mathbf r}, t+\Delta t)$, $\phi({\mathbf r}, t+\Delta t)$ and $\bar{\mathbf v}({\mathbf r}, t+\Delta t)$ of the forward time step satisfying eq.~\eqref{eq:dyn} and the boundary conditions ~\eqref{eq:bc0} through eq.~\eqref{bcje};
(iv)~Repeat the iteration scheme of the previous second and third steps till a final time, $t_{\rm final}$.

\subsection{Two-dimensional model for thin BCP films \label{sec:model-2DCH}}

We consider the limit of no macroscopic convection~\cite{DoiOnuki1992, Milner1993} and focus on thin films of concentrated BCP solutions (with relatively small $\phi$). Then from eq.~\eqref{eq:Matrix}, we have $M_{\rm S\psi}=M_{\psi \rm S}\approx 0$. That is, the cross-coupling between solvent and copolymer dynamics can be neglected. In this case, the above set of equations~\eqref{eq:dyn}-\eqref{eq:bch}, and \eqref{bcje} reduces to ~\cite{Goldenfeld1989,Muller2015}:
\begin{align} \label{eq:DynSyst}
{\partial_t  \phi}  &=- \nabla\cdot {\mathbf j}_{\rm S}, \quad {\mathbf j}_{\rm S}=- M_{\rm SS} \nabla \mu_{\rm S}\nonumber \\
{\partial_t \psi} &=- \nabla \cdot {\mathbf j}_{\psi}, \quad {\mathbf j}_{\psi}=-M_{\psi\psi} \nabla \mu_{\psi},
\end{align}
supplemented with boundary conditions at the bottom substrate, $z=0$,
\begin{align}\label{eq:DynSyst-bc0}
&\hat{\mathbf z} \cdot{\mathbf j}_{\rm S} =\hat{\mathbf z} \cdot{\mathbf j}_{\psi}=0, \nonumber \\
&\partial_z \psi=0, \nonumber \\
&\partial_z (\nabla^2\psi+q^2 \psi)=0
\end{align}
and at the top free surface, $z=h(t)$,
\begin{align}\label{eq:DynSyst-bch}
&\hat{\mathbf z} \cdot {\mathbf j}_{\rm S} = (1-\phi)j_{\rm e}/\rho_0, \nonumber \\
&\hat{\mathbf z} \cdot {\mathbf j}_{\psi} = -\psi j_{\rm e}/\rho_0, \nonumber \\
&\nabla^2\psi+q^2 \psi=0, \nonumber \\
&\partial_z (\nabla^2\psi+q^2 \psi)=0,
\end{align}
with the evaporation flux given by
\begin{align}\label{eq:DynSyst-je}
j_{\rm e}=\rho_0 v_{\rm e}(\phi-\phi^{\rm eq}).
\end{align}
For small volume fraction of solvent with $\mu_{\rm S} \approx 2 A_2 \phi$ and $A_2$ being the second virial coefficient, the dynamic equation for solvent in eq.~(\ref{eq:DynSyst}) reduces to simple diffusion equation with the collective diffusion constant given by $D_{\rm S} = 2 A_2 M_{\rm SS} \sim 2A_2 \zeta_{\rm S}^{-1}\phi^2$.

Furthermore, we take the limit that the relaxation dynamics throughout the film thickness ($z$-direction) is much faster than the lateral one ($xy$-plane), as is schematically shown in Fig.~\ref{Fig:Schematic}a. The variations of $\psi$ and $\phi$ are then negligible in the normal direction ($z$-axis), leading to $\psi=\psi(x,y,t)$ and  $\phi=\phi(t)$, and the changes of BCP morphology occur in the two dimensional $xy$-plane. The solvent volume fraction $\phi$ is a spatially uniform ``external" control parameter, which is determined by the evaporation rate of solvent at the free surface.

In the simplified limit, the temporal evolution of phase structure is determined by the following 2D Cahn-Hilliard equation~\cite{Goldenfeld1989,Muller2015}:
\begin{align} \label{eq:DynSyst2}
{\partial_t \psi} = \nabla_{\parallel} \cdot \left(M_{\psi\psi} \nabla_{\parallel} \mu_{\psi} \right).
\end{align}
in which $\nabla_{\parallel}$ denotes the 2D spatial gradient in the $xy$-plane. Moreover, from eqs.~\eqref{eq:DynSyst} to \eqref{eq:DynSyst-je}, we obtain the temporal evolution of the film thickness and the integrated solvent fraction, respectively, as
\begin{align}
\label{eq:ht2}
&\partial_th +j_{\rm e}/\rho_0=0.
\end{align}
\begin{align} \label{eq:DynSysthphis}
&{\partial_t (h \phi)} +j_{\rm e}/\rho_0 = 0,
\end{align}
respectively, from which we derive
\begin{align} \label{eq:htphist2}
&h=h_0/(1-\phi), \nonumber \\
&\partial_t \phi = - \tau_{\rm e}^{-1} (1-\phi)^2 (\phi-\phi^{\rm eq}).
\end{align}
That is, the morphological dynamics of thin BCP film in the $xy$-plane is described by eq.~\eqref{eq:DynSyst2} for the order parameter $\psi$, which in turn is controlled by the solvent fraction $\phi$ following eq.~(\ref{eq:htphist2}).

Finally, we would like to point out that the above limit is justified when the viscous time scale $\tau_{\rm v}= h_0^2/\nu$ (with $\nu$ being the kinematic viscosity of the BCP solution) is much smaller than the evaporative time $\tau_{\rm e} = h_0/v_{\rm e}$ for an initially stationary film to evaporate to dry film, {\it i.e.,} $\tau_{\rm v}\ll \tau_{\rm e} $. Furthermore, we focus on the simple case that the evaporation of solvent is \emph{slow} compared with vertical solvent/polymer diffusion, {\it i.e.,} $\tau_{\rm e} \gg \tau_{\rm S},\, \tau_{\rm P}$, with $\tau_{\rm S}= h_0^2/D_{\rm S}$ and $\tau_{\rm P}= h_0^2/D_{\rm P}$ being the diffusion times through the film thickness of the solvent and the polymer segments, respectively, and $D_{\rm S}$ and $D_{\rm P}$ the corresponding collective diffusion constants.

\subsection{Numerical Method and System Setup}

Using the classical central finite-difference method~\cite{Glotzer1995,Buxton2007},  we numerically integrate the dynamical equation (\ref{eq:DynSyst2}) for the order parameter $\psi$, and eq.~(\ref{eq:htphist2}) for $\phi$ (in the $xy$-plane), as is schematically shown in Fig.~\ref{Fig:Schematic}a. The length is measured in units of
\begin{equation}\label{eq:ell0}
\ell_0\equiv (\kappa/u)^{1/4},
\end{equation}
and from eq.~(\ref{eq:Fbsuba}) we obtain $\ell_0\approx 0.25 \lambda_0$.
Note that the number of lamellar period in our simulation box equals to an integer that is the closest to $L_{\rm x}/\lambda_0$. It is \emph{five} for $L_{\rm x}=20 \ell_0$ and dry lamellar spacing $\lambda_0 \approx 4.0 \ell_0$, as can be seen in Fig.~\ref{Fig:Schematic}c.

Time is measured in units of
\begin{equation}\label{eq:t0}
\tau_0 \equiv \ell_0^2/uM_{\psi\psi},
\end{equation}
which characterizes the time for copolymer reorganization in the scale of lamellar period $\lambda_0 \sim \ell_0$.

We employ periodic boundary conditions in the simulations to realistically model experimental setups of BCP films. The experimental lateral size is extremely large and cannot be modeled in any finite simulation box. However, periodic boundary conditions are the best way to model large systems as any effect of an artificially imposed lateral wall is removed. The box lateral size is set to include an integer number of lamellae in order to remove the artificial confinement effect that do not exist in the experiments. In addition, we have also varied the time step and mesh size to confirm the accuracy and efficiency of our numerical scheme. For mesh size of $0.05 \ell_0$, the simulations are stable for time steps smaller than $5 \times 10^{-8} \tau_0$, while for meshes of size $0.4\ell_0$, the simulations are stable for time steps smaller than $5\times 10^{-7} \tau_0$. We note that the results are not found to be sensitive to the mesh size and time step in the tested range. Specifically, in most of the presented simulations (except for Fig.~\ref{Fig:Finitesize1} and Fig.~\ref{Fig:Finitesize2}), we have chosen the mesh size to be $0.1 \ell_0$ and the time step $10^{-8}\tau_0$, for simulation box size of $L_{\rm x}=L_{\rm y}=20 \ell_0$.

We carry out a numerical study for the effects of solvent swelling and solvent evaporation on the morphology of a dry BCP film with a specific defect as shown in Fig.~\ref{Fig:Schematic}c. In our simulations, the initial defective morphology of the dry film is prepared at $N\chi=30$, starting from an artificially designed initial morphology that is close to the desired defect structure as shown in Fig.~\ref{Fig:Schematic}c. In addition, when the dry film is exposed to solvent in experiments, the uptake (swelling) rate of solvent in most cases strongly depends on the morphology of BCP film even for a non-selective solvent~\cite{Zhang2014}. However, the exposure time to solvent vapor (the annealing time) is usually longer than that required for the film to swell and reach an equilibrium (solvent) volume fraction~\cite{Gu2014a,Zhang2014}. In our simulations, we ignore the effect of solvent uptake process and assume ``instantaneous" saturation to equilibrium (solvent) volume fraction $\phi^{\rm eq}$ as schematically shown in Fig.~\ref{Fig:Schematic}b. After an annealing time $\tau_{\rm a}$, the solvent is removed by a controlled evaporation rate, $1/\tau_{\rm e}$, according to eq.~(\ref{eq:htphist2}).

\section{Results and Discussion \label{sec:results}}

We present results on the effects of two key parameters on defect removal during SVA: solvent-swelling ratio ${\cal R}$, that is a monotonic function of equilibrium volume fraction $\phi^{\rm eq}$, and the normalized solvent-evaporation rate, $\alpha_{\rm e}= \tau_0/\tau_{\rm e}$.

\subsection{Effects of solvent-swelling ratio \label{sec:results-swelling}}

The swelling of the BCP film is one-dimensional along the normal $z$-direction, due to the lateral confinement of fixed area set by the in-plane periodic boundary conditions, as shown in Fig.~\ref{Fig:Schematic}a. The swelling ratio ${\mathcal R}\equiv V/V_0$ is then given by
\begin{equation}\label{eq:R}
{\mathcal R}=h/h_0=\frac{1}{1-\phi^{\rm eq}}
\end{equation}
from the conservation of copolymers and eq.~\eqref{eq:htphist2} with $h_0$ and $h$ being the film thickness of the original dry film and the swollen film, respectively.

\emph{Negligible changes in lamellar spacing with $\beta =0$}. As pointed out in Section~II, there are two major effects of a non-selective solvent that swells a BCP film: dilution effect acting on $\chi_{\rm eff}(\phi)$, eq.~(\ref{eq:chieff}), and increase of the lamellar spacing $\lambda(\phi)$, eq.~(\ref{eq:spacing1}).
We first concentrate on a simpler case for which the solvent-induced change in lamellar spacing, $\lambda$, is negligible during solvent annealing. Hence,  $\lambda \approx {\rm const.}$. This is modeled by the limit $\beta \approx 0$ in eq.~(\ref{eq:spacing1}). It can result from the weak dependence of either $\chi_{\rm eff}$ on $\phi$ or $\lambda$ on the incompatibility $N\chi_{\rm eff}$~\cite{Gu2014a,Gu2014b}.

\begin{figure}[htbp]
\centering
  \includegraphics[clip=true, viewport=5 5 680 460, keepaspectratio, width=0.4\textwidth]{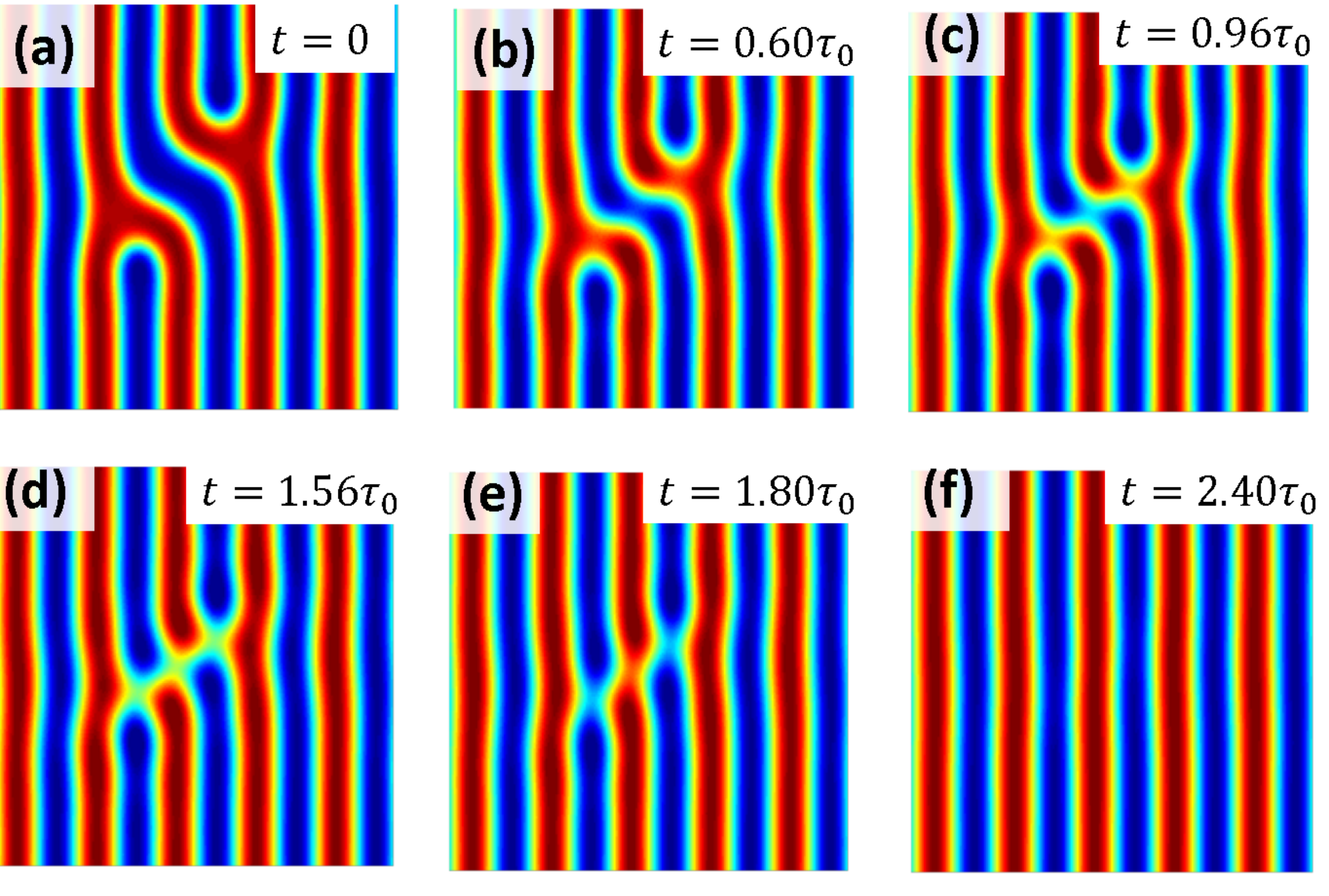}
  \caption {Defect removal upon instantaneous solvent swelling for $\beta=0$ leading to $\lambda = {\rm const.}$ only due to dilution effect on the $\chi_{\rm eff}(\phi)$ parameter, given the term $\chi (1-\phi)$ in eq.~(\ref{eq:chieff}).  Other parameters are $N\chi =30$ and $\phi^{\rm eq}=0.6$ that corresponds to a swelling ratio of ${\cal R}=2.5$.
  } \label{Fig:Dilution}
\end{figure}

From the simulations we can infer that the defect stability strongly depends on $\phi^{\rm eq}$ (or swelling ratio $\cal R$ given in eq.~(\ref{eq:R})) in the swollen film.
For a relatively small ${\mathcal R}$ (small $\phi^{\rm eq}$), the system is kinetically trapped (metastable). On the other hand, in the limit of very large swelling ratio, occuring for $\chi_{\rm eff}<\chi_{\rm c}$ (equivalently, for ${\cal R}>2.86$ or $\phi^{\rm eq}>0.65$), the BCP undergoes an order-disorder transition (ODT), and the BCP solution evolves into a disordered phase (no lamellar phase). As shown in Fig.~\ref{Fig:Dilution}, only in a narrow window of swelling ratio, $2.5<{\cal R}<2.86$ (or equivalently $0.6<\phi^{\rm eq}<0.65$), the copolymer segments gain large enough mobility and maintain an ordered structure such that a defect-free lamellar structure is obtained. Note that this window in ${\mathcal R}$ values should be considered only for the case of very long annealing times and very slow solvent evaporation. In other cases, it can be varied significantly by changing the solvent evaporation rate, $\alpha_e$, as shown in the inset of Fig.~\ref{Fig:diagram}.

The mechanism of defect removal by added solvent  (reducing $\chi$) is very similar to thermal annealing (TA), where $\chi \sim 1/T$~\cite{Li2015,Li2014,Hur2015,Muller2016,Muller2018}. The presence of a threshold value of incompatibility in $N\chi_{\rm eff}$ (given by a threshold of $\phi$ or $\cal R$) in SVA has also been seen in TA~\cite{Li2015,Li2014,Hur2015,Hur2018,Song2018,Muller2016,Muller2018}, where $N\chi$ is a function of $T$. For small $N\chi_{\rm eff}$ (large swelling ratio), defects are unstable and can be removed spontaneously as there are no energy barriers to overcome. For large $N\chi_{\rm eff}$ (small swelling ratio), defects are metastable and some free-energy barriers need to be overcome. However in the simulation method, the morphology dynamics is driven by free-energy minimization. Therefore, we can neither examine the stability of a metastable defect nor quantify the dependence of the energy barrier on $N\chi_{\rm eff}$ or $\phi$. We can only check when the energy barrier of the defect removal vanishes and the defect becomes unstable as ${\mathcal R}$ is increased (see discussions above), as well as examine the dynamic processes of spontaneous defect removal.

\begin{figure}[htbp]
  \centering
  \includegraphics[clip=true, viewport=1 1 750 180, keepaspectratio, width=0.45\textwidth]{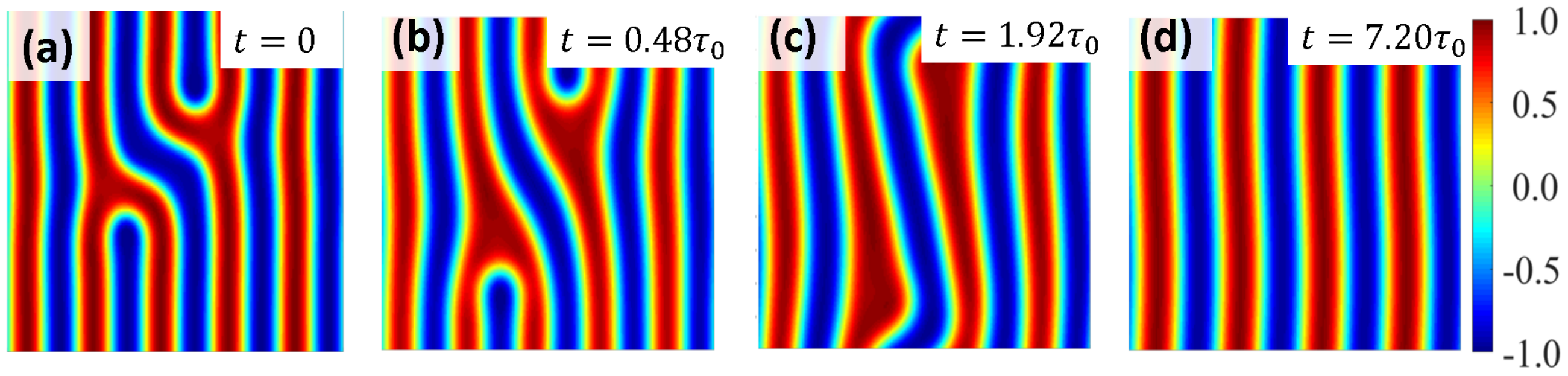}
  \caption {Destabilization of defects upon instantaneous solvent swelling due to an \emph{increase} in the lamellar spacing. Here, we take $N\chi=30$, $\phi^{\rm eq}=0.4$ (corresponding to ${\cal R}\approx 1.67$) and $\beta=-1.0$ as defined in eq.~(\ref{eq:spacing1}).
  } \label{Fig:Spacing}
\end{figure}

Such dynamic processes have a symmetric pathway in the sense that the two horizontal arms break simultaneously, as seen in Fig.~\ref{Fig:Dilution}a. This is in contrast to the kinetic pathway of defect removal that is seen in TA studies. Here, the defect removal pathway is asymmetric~\cite{Li2014} in the sense that the two defect horizontal arms (shown in Fig.~\ref{Fig:Schematic}c) break sequentially, one after the other, by crossing two consecutive free-energy barriers~\cite{Hur2018,Muller2018,Song2018}.

\emph{Solvent-induced increase in lamellar spacing with $\beta<0$}. We also find on general ground that the change in lamellar spacing $\lambda$ during solvent annealing of BCP films is visible and cannot be neglected~\cite{Sinturel2013,Gu2014a,Gu2016}.
At small $\phi^{\rm eq}$ below some critical value, $\phi^{*}$~\cite{Shibayama1983a,Shibayama1983b}, the spacing $\lambda$ increases with increasing $\phi$. In Fig.~\ref{Fig:Spacing}, we take $\phi^{\rm eq}=0.4$, which is smaller than $\phi^{*}$. Also $\beta <0$, the equilibrium lamellar spacing increases from $\lambda_0\approx 4 \ell_0$ for the dry state to $\lambda\approx 6\ell_0$ for the swollen state. Correspondingly, the energy-preferred number of lamellar periods, given by $L_x/\lambda_0$, changes from five to three. Figure~\ref{Fig:Spacing} shows that an increase in lamellar spacing tends to drive long-wavelength structural evolution. After defect removal is completed, the system evolves into an intermediate metastable defect-free state with \emph{four} lamellar periods instead of energy-preferred three periods. This result again indicates that the addition of solvent alters the kinetic pathway of the microstructure evolution of the BCP film.

\begin{figure}[htbp]
\centering
  \includegraphics[clip=true, viewport=1 1 750 180, keepaspectratio, width=0.45\textwidth]{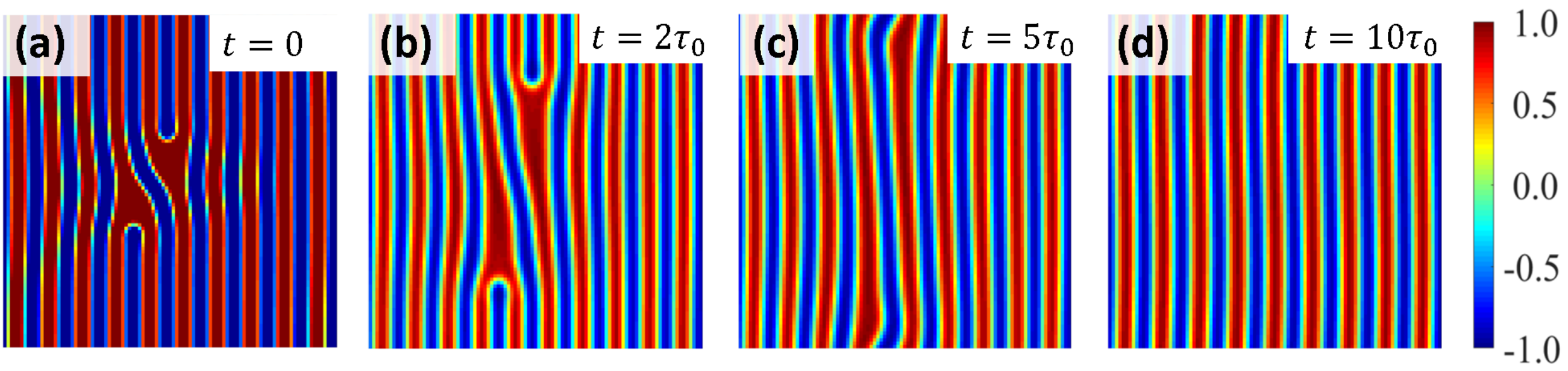}
  \caption {Defect removal in a film with larger lateral dimensions,  $L_{\rm x}=L_{\rm y}=40 \ell_0$. The mesh size is $0.4 \ell_0$ and the time step is $10^{-7}\tau_0$. All other parameter values are the same as in Fig.~\ref{Fig:Spacing}.
  } \label{Fig:Finitesize1}
\end{figure}

Comparing the results obtained with spacing changes ($\lambda \ne {\rm const.}$) as in Figs.~\ref{Fig:Spacing} and \ref{Fig:Finitesize1} to those without spacing changes ($\lambda={\rm const.}$ as in Fig.~\ref{Fig:Dilution}), we find that in both cases the defect-free state can be obtained only in a narrow window of $\phi$. This window is large enough to make defect state unstable as well as it exists below ODT to be in the ordered lamellar phase. This observation agrees with experiments~\cite{Baruth2014}. Even more interestingly is the fact that the defect structure of a BCP solution with $\phi^{\rm eq}=0.4$ is still metastable in the latter case without changes in $\lambda$. However, the domain becomes unstable in the former case with spacing increases (as shown in Fig.~\ref{Fig:Spacing}). We, therefore, assert that the solvent-induced increase in $\lambda(\phi)$ facilitates a more efficient removal of defects. It seems reasonable to conjecture at this point that this important observation is more general than the specific model we have presented in this study.

\begin{figure}[htbp]
  \centering
  \includegraphics[clip=true, viewport=1 1 750 500, width=0.4\textwidth]{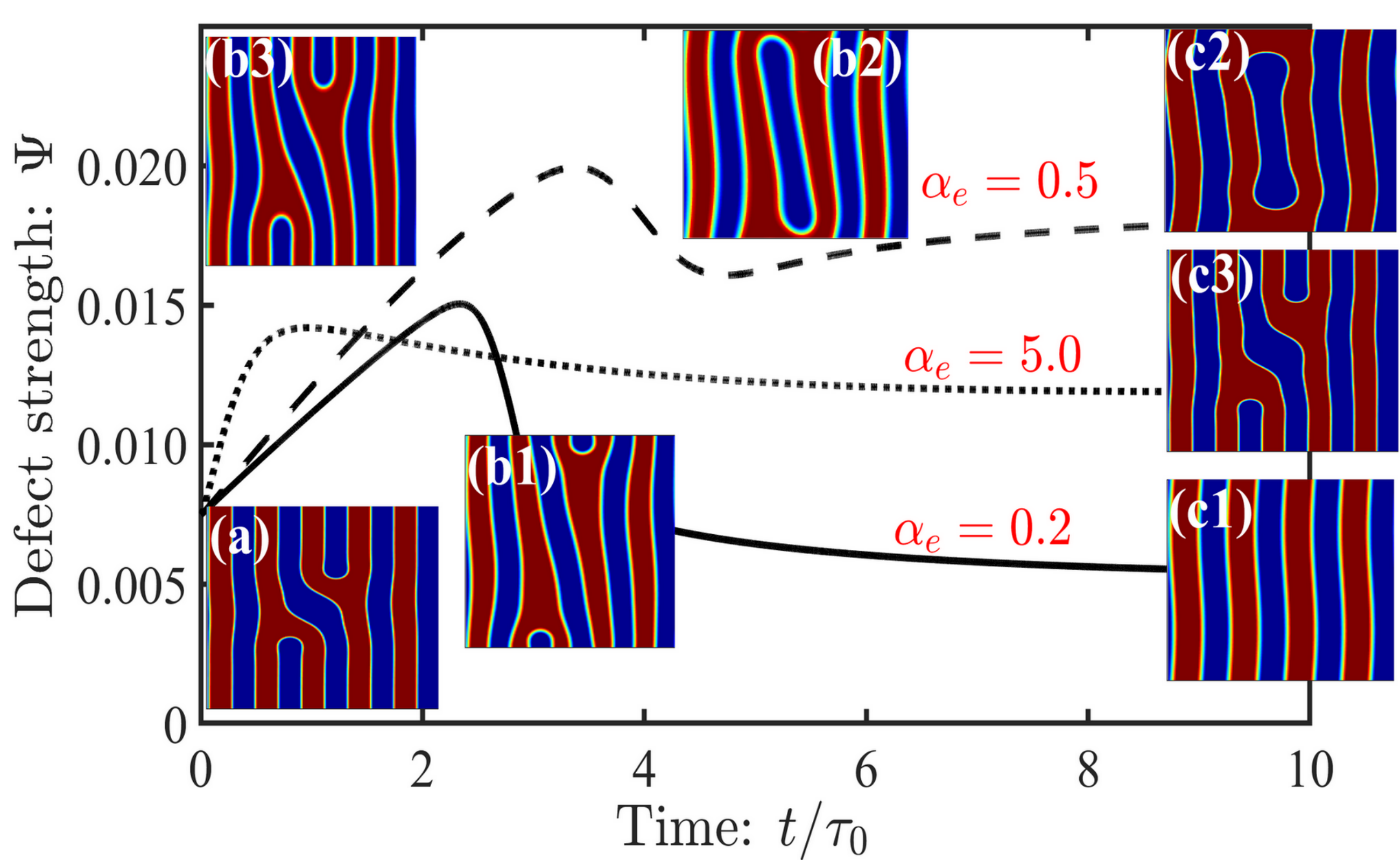} 
  \caption {The temporal evolution of the defect structure (shown in Fig.~\ref{Fig:Schematic}c) of a swollen BCP film at various evaporation rates $\alpha_{\rm e}= 0.2, \, 0.5, \, 5.0$ after annealing time $\tau_{\rm a}=0.48 \tau_0$. The Flory parameter is $N\chi=30$, $\phi^{\rm eq}=0.4$, corresponding to ${\cal R}\approx 1.67$  and $\beta=-1.0$ as in eq.~(\ref{eq:spacing1}). The \emph{defect strength} is defined by $\Psi\equiv  \sum_{i,j}(\psi_{i,j}-\psi_{i,1})^2$ with $i, \,j$ denoting the sites, and is a convenient way to quantify the process of defect removal in the computational box.
  Snapshots of the temporal evolution of the original defect (a) (no solvent) are shown for different evaporation rates, $\alpha_{\rm e}$. For slow evaporation, the defect (a) is removed and the system evolves spontaneously towards a defect-free state (c1), after crossing an intermediate transition state (b1).  In contrast, for faster evaporation, the defect (a) is either kinetically trapped as in state (c3) after crossing the intermediate transition state (b3), or becomes unstable and is re-trapped in a new defect state (c2) after crossing (b2).
  } \label{Fig:evaporation}
\end{figure}

\emph{Effects of the system size for $\beta<0$}. Lateral confinement set by periodic boundary conditions also plays a role in determining the final morphology of the thin film. We have carried out larger-scale simulations as shown in Fig.~\ref{Fig:Finitesize1}, where the system sizes are doubled in both lateral directions, (see Fig.~\ref{Fig:Spacing}).
Comparing these results in Fig.~\ref{Fig:Finitesize1} with those of Fig.~\ref{Fig:Spacing}, we find that for larger systems, it takes longer times to remove the defect, because the distance of the defect from the boundary is larger. However, the dynamics of defect removal is qualitatively similar for the two different system sizes. In both cases, the defects are unstable and move along the lamellae when the same fraction of solvent is added.

\subsection{Effects of solvent-evaporation rate \label{sec:results-evaporation}}

It is well-known that the final morphology of the dried BCP film depends, not only on the swelling ratio $\cal R$ but also on the solvent evaporation rate, $\alpha_{\rm e}$ ~\cite{Sinturel2013,Gu2014a,Zhang2014,Gu2016}. To examine this dependence, we carry out simulations for swollen films with different swelling ratio $\cal R$ ({\it i.e.,} with different $\phi^{\rm eq}$) for evaporation at different $\alpha_{\rm e}$ rates. As schematically shown in Fig.~\ref{Fig:Schematic}b, the solvent is removed by evaporation following eq.~(\ref{eq:htphist2}) after annealing time, $\tau_{\rm a}$.

\emph{Dependence of dried BCP film morphology on $\alpha_e$ at different $\cal R$.} For small swelling ratio ${\cal R} <1.2$ ($\phi^{\rm eq} < 0.15$), the amount of solvent in the swollen film is not sufficient to enhance polymer chain mobility and the original defect remains metastable and kinetically trapped. After solvent evaporation, the original defect persists in the dried film and is found to be independent of the evaporation rate.

In contrast, at large swelling ratio ${\cal R} >2.86$ ($\phi^{\rm eq}> 0.65$) where $N\chi_{\rm eff} < N\chi_{\rm c}$, the system passes through the ODT and the swollen BCP film is disordered. After the solvent evaporates, the film re-enters into the ordered phase but the final dry film is usually trapped in a poorly ordered state characterized by a significant amount of defects. Furthermore, the final phase morphology strongly depends on the evaporation rate. To obtain a defect-free state, one has to direct the film morphology by controlling the solvent evaporation rate~\cite{Albert2011,Fredrickson2013,Fredrickson2016}, or by applying a shear flow~\cite{Marencic2007,Shelton2017}.

At intermediate swelling ratio ${\cal R} \sim 1.7$ ($\phi^{\rm eq} \sim  0.4$), the copolymer chains have sufficient mobility and the defect is removed spontaneously if the annealing time $\tau_{\rm a}$ is long enough, as shown in Fig.~\ref{Fig:Spacing}. In SVA experiments, the solvent evaporates after some small annealing time, $\tau_{\rm a}<\tau_0$. This motivated us to choose an annealing time $\tau_{\rm a}=0.48 \tau_0$, after which the solvent is evaporated with controlled rate, $\tau_{\rm e}^{-1}$.

\emph{Quantifying the defect strength of the BCP film at different $\alpha_e$}. In order to quantify how the evaporation rate influence the defect dynamics, we introduce for convenience a new parameter defined as $\Psi\equiv  \sum_{i,j}(\psi_{i,j}-\psi_{i,1})^2$, with $i\,,j$ denoting the sites in the computational box. Although this ``defect strength'' parameter is not uniquely defined for a particular morphology, and even is not meaningful if tilting of the lamellae occurs, it is simpler to analyze $\Psi$ than to look directly at the specific film morphology. From the temporal evolution of this parameter shown in Fig.~\ref{Fig:evaporation}, we can quickly see that the phase structure of the dried BCP film strongly depends on the evaporation rate $\alpha_{\rm e}=\tau_0/\tau_{\rm e}$. The initial defect (Fig.~\ref{Fig:evaporation}(a)) can be removed only via \emph{slow} evaporation with $\alpha_{\rm e} \ll 1$, as in Fig.~\ref{Fig:evaporation}(c1) (a snapshot before the defect is removed can be seen in Fig.~\ref{Fig:evaporation}(b1)). At intermediate evaporation rates with $\alpha_{\rm e}\sim 1$, the initial defect becomes unstable but is re-trapped to another defective structure (but with larger spacing) as shown in Fig.~\ref{Fig:evaporation}(c2), while at fast evaporation, $\alpha_{\rm e} \gg 1$, the defect persists as shown in Fig.~\ref{Fig:evaporation}(c3).

It is interesting to note that in the simulations, we have not seen any annihilation of the new intermediate defect (Fig.~\ref{Fig:evaporation}(c2)) up to times $\sim 620 \tau_0$. However, such a defect has been found in previous self-consistent field simulations~\cite{Muller2016} to be thermodynamically unstable, but with very long annihilation times. The understanding of this inconsistency worth further studies, although it will be difficult to perform using our method. Other methods such as the string method can be used in future studies in order to explore the energy landscape.

\begin{figure}[htbp]
  \centering
  \includegraphics[clip=true, viewport=1 1 750 180, keepaspectratio, width=0.45\textwidth]{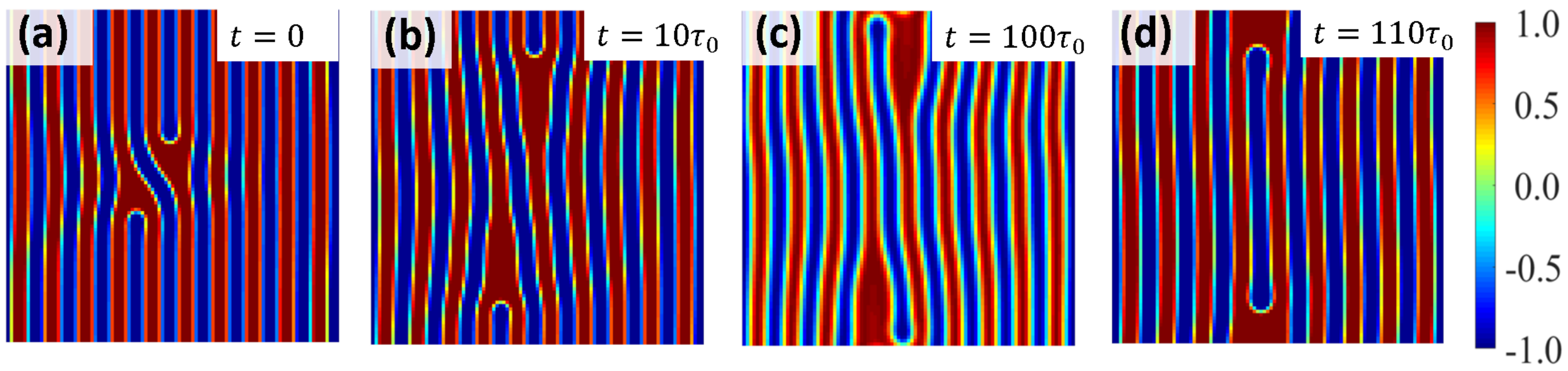}
  \caption {Defect destabilization and its re-trapping into an intermediate defect state in a film with larger lateral dimensions, of $L_{\rm x}=L_{\rm y}=40 \ell_0$. The mesh size is $0.4 \ell_0$ and time step is $10^{-7}\tau_0$. All other parameters are the same as in Fig.~\ref{Fig:evaporation}(b2) and (c2) with evaporation rate $\alpha_{\rm e}=0.5$.
  } \label{Fig:Finitesize2}
\end{figure}

\emph{Influence of finite system size}. To investigate the influence of the finite system size on the evaporation rate results (Fig.~\ref{Fig:evaporation}), we have carried out larger-scale simulations and present them in Fig.~\ref{Fig:Finitesize2}. The system size is doubled in both lateral directions, and the simulations are applied to the specific case presented in Fig.~\ref{Fig:evaporation}(c2) with evaporation rate $\alpha_{\rm e}=0.5$.
Comparing these results, we find that for larger systems, it takes a longer time for the defect to be removed as can be seen in Fig.~\ref{Fig:Finitesize1}, because of the larger distance of the defect from the boundary. However, the dynamics of the defect removal remains qualitatively similar. For an intermediate solvent evaporation rate ($\alpha_{\rm e}=0.5$), the initial defect becomes unstable and the thin film is re-trapped into a new intermediate-defect state, as shown in Figs.~\ref{Fig:evaporation}(c2) and \ref{Fig:Finitesize2}.

\emph{Morphology of dried film emerging  from the competitions in multiple time scales}. All these results can be understood in terms of the total sum of annealing time $\tau_{\rm a}$ and the evaporation time $\tau_{\rm e}$. The combined $\tau_{\rm a}+\tau_{\rm e}$ gives the time allowing the copolymers to re-organize during SVA (with characteristic time $\tau_0$). For a given small $\tau_{\rm a}$, when the solvent is removed rapidly, the copolymer loses its mobility almost instantly, and is kinetically trapped into the defect-swollen morphology.
When the solvent is removed gradually, on the other hand, the lamellar spacing changes and the defects are destabilized. Moreover, the BCP microdomains have time to relax so that the defects can be removed, just as in the case of very long annealing times without solvent evaporation. In the annealing process of thin BCP films on neutral substrates by non-selective solvents, the most efficient pathway to remove defects is first to swell the BCP film sufficiently such that defects become unstable, and only then to evaporate the solvent rapidly when the defect is almost ``healed" spontaneously.

\begin{figure}[htbp]
  \centering
  \includegraphics[clip=true, viewport=1 1 750 600, keepaspectratio, width=0.4\textwidth]{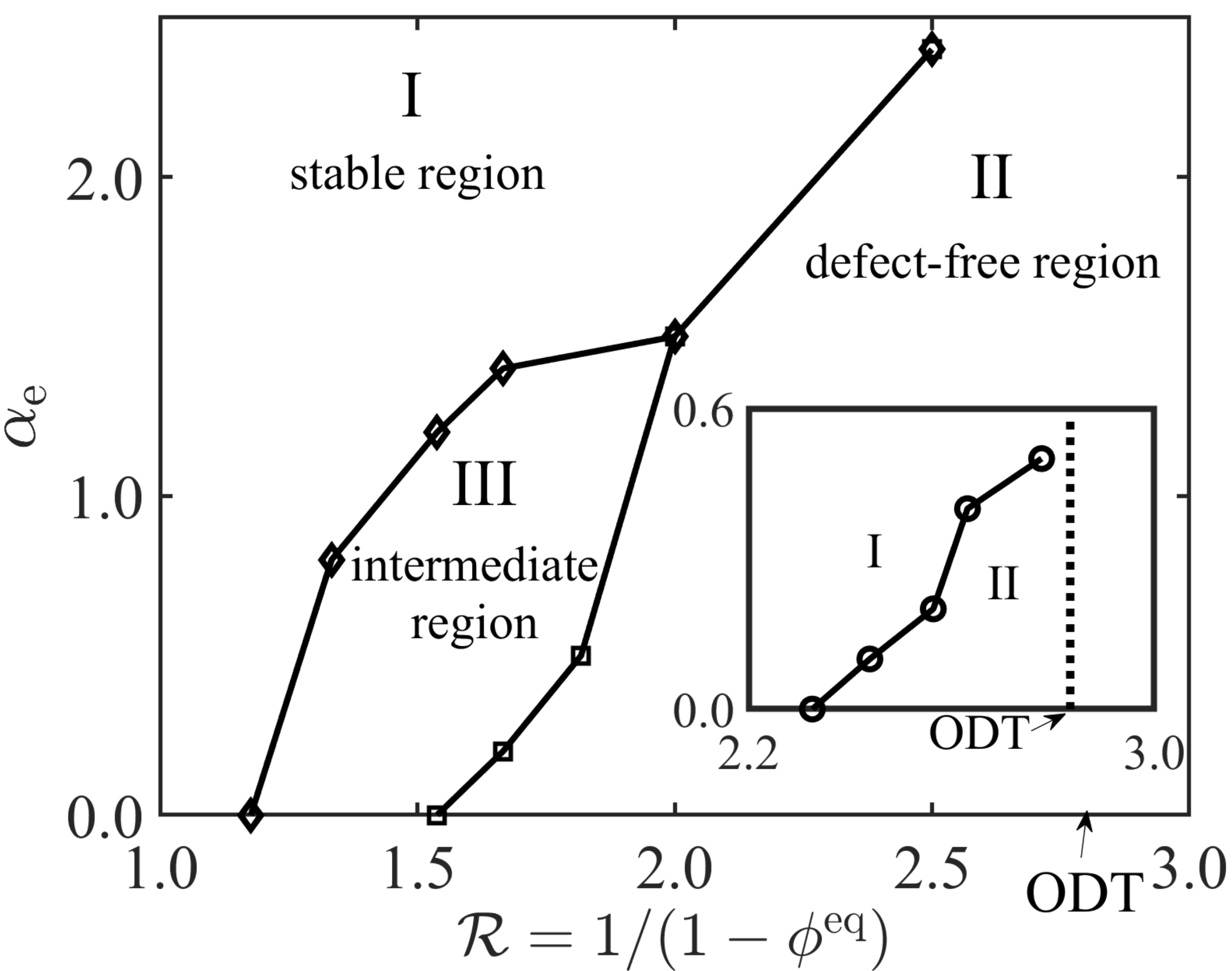}
  \caption {Defect stability diagram shown in Fig.~\ref{Fig:Schematic}c during SVA due to the dilution effect and changes in lamellar spacing. The diagram is plotted in terms of two parameters: the swelling ratio ${\cal R} = 1/(1-\phi^{\rm eq})$ and the evaporation rate
  $\alpha_{\rm e}=\tau_0/\tau_{\rm e}$. We set $N\chi=30$ and the solvent-induced spacing exponent $\beta=-1.0$ as in eq.~(\ref{eq:spacing1}). The solvent is evaporated after an annealing time $\tau_{\rm a}=0.48 \tau_0$. Three regions are identified: \emph{Stable region} (I), where the original defect is still (meta-) stable; \emph{Defect-free region} (II), where  the defect becomes unstable and is removed spontaneously; \emph{Intermediate region} (III), where the original defect is unstable but is re-trapped into a new defect state. The lines are shown to guide the eyes.  (Inset) Same defect stability diagram but with $\lambda={\rm const.}$ (no spacing changes are allowed) during SVA.
  }
  \label{Fig:diagram}
\end{figure}

\emph{Defect stability diagram showing the dependence of the dried film morphology on $\cal R$ and $\alpha_{\rm e}$}. Finally, we note that the dependence of the final morphology of dried BCP films on solvent-swelling ratio $\cal R$ and solvent-evaporation rate $\alpha_{\rm e}$ can be summarized in a defect stability diagram, as shown in Fig.~\ref{Fig:diagram}. Three regions are identified: (I) Stable region at small $\cal R$ and large $\alpha_{\rm e}$, in which the original defect is still stable or metastable and persists in the dried film; (II) Defect-free region at large $\cal R$ and small $\alpha_{\rm e}$, in which the defect becomes unstable and is removed spontaneously; (III) Intermediate region, in which the original defect is unstable but can be re-trapped into a new defect state. Such trapping or transition into a imperfect lamellar structure with a new defect has been reported previously. Using the TA technique, inplane defect pairs~\cite{Hur2018,Muller2016,Muller2018} and cross-sectional out-of-plane defect pairs~\cite{Song2018} have been observed.

When the solvent-induced changes in lamellar spacing, $\lambda$, are negligible (see inset of Fig.~\ref{Fig:diagram} where $\lambda={\rm const.}$),  The stability diagram contains only two regions: (I) Stable-defect region; and, (II) Defect-free region. Note that the intermediate region (III) does not exist anymore. Moreover, the defect-free region (II) is much smaller when the solvent-induced spacing changes is taken into account. Another remark is that the numerical accuracy of the stability diagram depends on the choice of the mesh size. Only a small enough mesh size can fully resolve the ``shape" of the free-energy landscape and gives accurate stability. Here, we have chosen the mesh size to be $0.1\ell_0$ as defined in eq.~(\ref{eq:ell0}), and checked that it gives the same diagram as with a smaller mesh size twice as small, $0.05\ell_0$. Furthermore, the defect stability  also depends on the ratio of the lamellar spacing between swollen state and dry state, ${\cal R}$ and the exponent $\beta$ defined in eq.~(\ref{eq:spacing1}), as will be discussed in the next section.
Our computational prediction agrees, at least qualitatively, with experimental observations~\cite{Gu2014a,Gu2014b}, and indicates that the two key parameters ${\cal R}$ and $\alpha_{\rm e}$ not only affect the lamellar spacing of BCPs but also govern the removal of defects and lateral ordering of the BCP microdomains. However, the predicted stability diagram still needs to be verified quantitatively in future experiments and more extended simulations.

\subsection{Effects of spacing decrease induced by large $\phi$ \label{sec:spacingdecrease}}

As discussed in Sec.~\ref{sec:energy}, the effect of added solvent on the lamellar spacing can be divided into two regimes, separated by a crossover solvent volume fraction, $\phi^{*}$. In previous sections, we have taken $\phi^{*}$ to be large such that $\phi< \phi^{*}$ and the lamellar spacing increases with added solvent. Here, we choose small $\phi^{*}$ and explore the effect of spacing changes in the two regimes, and concentrate, in particular, on the effect that increasing $\phi$ has on decreasing $\lambda$.

\begin{figure}[htbp]
  \centering
  \includegraphics[clip=true, viewport=1 1 750 180, keepaspectratio, width=0.45\textwidth]{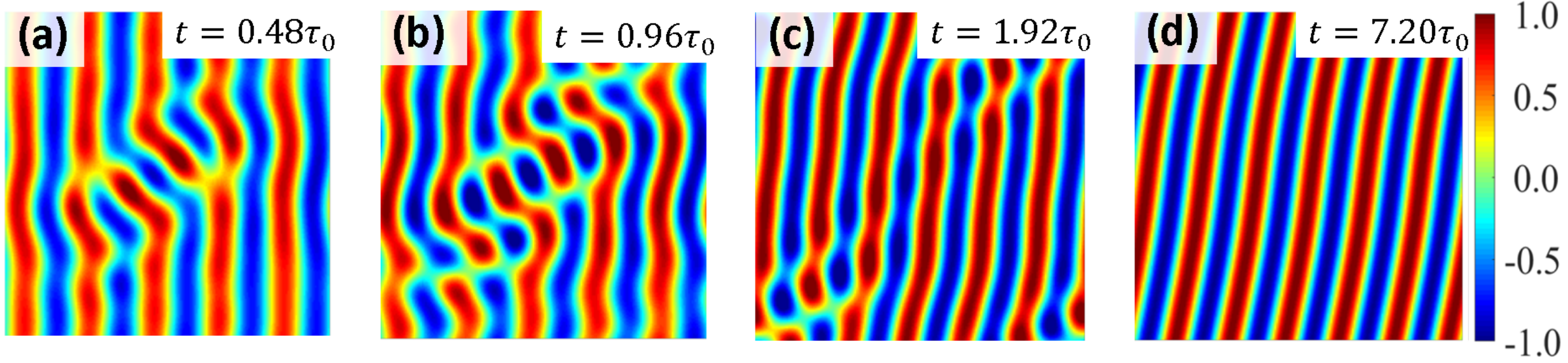}
  \caption {Destabilization of defects upon instantaneous solvent swelling due to the changes in the lamellar spacing $\lambda$. The used parameters are: $N\chi=30$, $\phi^{\rm eq}=0.6$ and $\phi^{*}=0.1$. As $\phi$ increases, $\lambda$ increases (with an exponent $\beta=-1.0$) when $\phi<\phi^{*}$, but decreases (with $\beta=1/3$) when $\phi>\phi^{*}$.
  } \label{Fig:Spacing2}
\end{figure}

\begin{figure}[htbp]
  \centering
  \includegraphics[clip=true, viewport=1 1 750 180, keepaspectratio, width=0.45\textwidth]{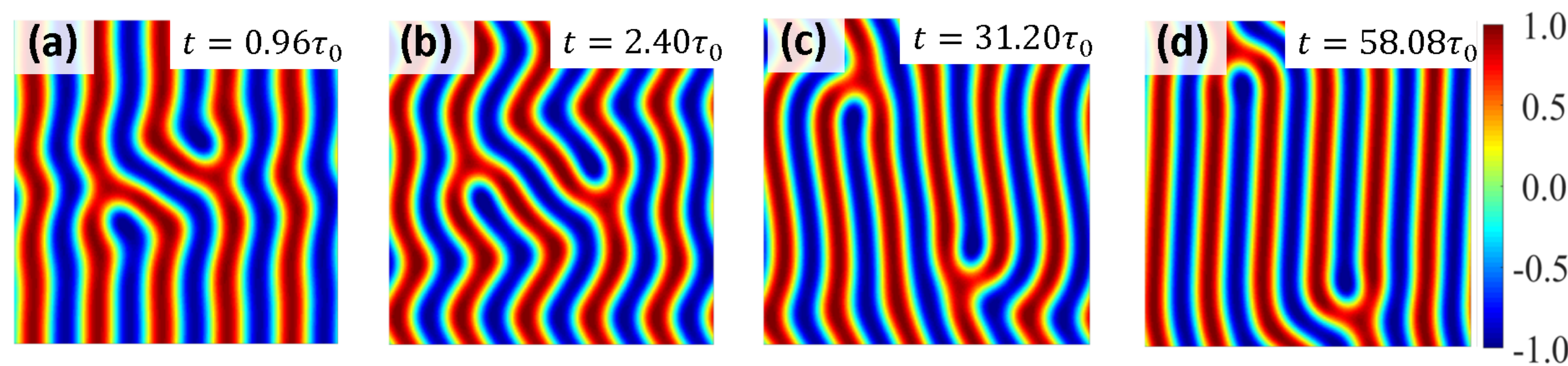}
  \caption {Destabilization of defects upon instantaneous solvent swelling due to the change of the lamellar spacing $\lambda$. The used parameters: $N\chi=30$, $\phi^{\rm eq}=0.4$, and $\beta=1/3$, as defined in eq.~(\ref{eq:spacing1}).
  } \label{Fig:Spacing1}
\end{figure}

In both regimes, $\phi>\phi^{*}$ and $\phi< \phi^{*}$ we find that the change in lamellar spacing, $\lambda$, destabilizes the defect, as shown in Figs.~\ref{Fig:Spacing1} and ~\ref{Fig:Spacing2}.
However, the kinetics of defect instability is very different in the two regimes. At small  $\phi< \phi^{*}$, the solvent-induced $\lambda$ increases and tends to drive long wavelength structural evolution, as shown in Fig. 3. Moreover, the kinetically-controlled spacing-increase (when $\phi<\phi^{*}$) during solvent annealing is essential for removing BCP defects efficiently, as shown in Fig.~\ref{Fig:Spacing2}. In contrast, at large  $\phi> \phi^{*}$, the solvent-induced $\lambda$ decreases  and introduces small wavelength undulations, as shown in Fig.~\ref{Fig:Spacing1}.  Note that the obtained defect-free lamellae adopt a tilted orientation in the $xy$-plane to adjust to the smaller domain size under the constraints imposed by a fixed $L_{\rm x} \times L_{\rm y}$ size of the simulation box.

%
We find that the solvent-induced $\lambda$ decrease at large $\phi$ induces small-wavelength undulations in contrast to the long wavelength structural evolution, as shown in Fig. ~\ref{Fig:Spacing}.  However, these small-wavelength undulations are usually not enough to remove defects as shown in Fig.~\ref{Fig:Spacing1}. Efficient defect removal occurs only for cases with significant spacing-increase during solvent annealing at small $\phi<\phi^{*}$.

\section{Conclusions \label{sec:conclusion}}

We have constructed a two-fluid model to study the microphase separation dynamics of block copolymer (BCP) thin films and applied it to solvent vapor annealing (SVA) processes. We focus on the effect of added (non-selective) solvent on the defect annihilation in ordered BCP thin films on neutral substrates. More specifically, we investigate the solvent-facilitated removal of a single defect occurring in symmetric lamellar BCP thin films with lamellae perpendicular to the substrate.

In our phenomenological two-fluid model, we have included two key features of the added solvent: weakening the unfavorable interactions between polymer segments, and changing the characteristic spacing of the BCP lamellae. We find that the solvent-induced increase of the lamellar spacing in dense polymer solution (namely, at small solvent volume fraction, $\phi$) facilitates an efficient removal of defects. The kinetic pathway of defect removal during SVA is quite different from that of thermal annealing (TA) and depends sensitively on the volume fraction, $\phi$, and on the dynamics of added solvent. The final morphology of the dried BCP film after the SVA process depends not only on solvent-swelling ratio ${\cal R}$ (or solvent volume fraction, $\phi$) but also on the solvent-evaporation rate, $\alpha_{\rm e}$. Such a dependence is summarized in a stability diagram (Fig.~\ref{Fig:diagram}), and can be verified in future experiments.

Based on these results, we are able to make an important observation about the annealing of thin BCP films  by non-selective solvents. The most efficient pathway to remove the defects is to first swell the film to a sufficiently large solvent fraction where defects become unstable. Then, as a second step, evaporating the solvent rapidly when the defect almost ``heals" spontaneously.

We would like to conclude with a few qualitative remarks on future directions and the relation of this work to experiments.
The competition between finite system size and lamellar spacing is essential in determining the dynamics of defect removal as well as the final morphology of the film. However, in this work, we have only chosen two system sizes for a given lamellar spacing. Systematic investigations of finite-size effect should be done in the future by gradually increasing the system size to cover more periods and to unify our predictions.

The swelling dynamics of BCP films strongly depends on the BCP morphology~\cite{Gu2014a,Zhang2014}. For example, films with perpendicular lamellae swell faster than parallel ones because a higher amount of solvent can penetrate through the A/B interface that is exposed to the free surface. This effect is associated with the inhomogeneous distribution of solvent within microphase separated BCP film. In our work, this is omitted because we have neglected the coupling of the swelling dynamics with BCP morphology dynamics. Nonetheless, the inhomogeneous solvent distribution and morphology-dependent swelling rate can have a significant effect on defect removal during solvent annealing process.

We have assumed that the relaxation dynamics throughout the film thickness is much faster than their lateral dynamics. Based on this assumption, we have neglected completely any variations of the structure and dynamics through the film thickness. However, it has been shown in many experiments~\cite{Albert2010,Sinturel2013,Gu2014a,Zhang2014} that the structure and dynamics thoughout the film thickness can be of importance during SVA processes. For example, as solvent is introduced or removed, there must be a rearrangement of the BCP chains, and some tilting of the lamellae is expected.
Moreover, film thickness is an important parameter, used as a mean allowing a direct access to the dynamics of the structural rearrangements~\cite{KimLibera1998,Cavicchi2007,Albert2011,Hao2017}. By controlling the film thickness in the fully swollen state, Cavicchi et al.~\cite{Cavicchi2007} found for cylindrical BCP phase that either parallel or perpendicular orientation could be obtained. In addition, Kim and Libera~\cite{KimLibera1998} suggested that solvent diffusion towards the free surface and solvent concentration gradients within the film imposed by different solvent evaporation rates are responsible for the different microdomains orientations.

More recently, experiments ~\cite{Albert2011} indicated that for a swollen film of parallel cylinder morphology, when the solvent is removed instantaneously, well-ordered parallel cylinders occur. However, if the solvent is removed very slowly, the parallel cylinders reorient and become perpendicular to the substrate. Finally, it has been known that surface tension that was neglected in this work can also play some role in determining the orientation of the lamellar BCP film~\cite{Sinturel2013,Hao2017}.

We believe that the insights provided in this theoretical work on defect removal from lamellar thin BCP films by SVA has the potential to greatly expand and diversify the range of applications of solvent-assisted, directed self-assembly of BCP thin films.

\vspace{12pt}
\noindent \textbf{Acknowledgments}
 \vspace{6pt}

\noindent We would like to thank S. Komura and T. Qian for discussions and helpful suggestions. This work was supported in part by Grant No. 21822302 and 21434001 of the National Natural Science Foundation of China (NSFC), the joint NSFC-ISF Research Program, jointly funded by the NSFC under Grant No. 51561145002 and the Israel Science Foundation (ISF) under Grant No. 885/15. X. X. was supported by Guangdong Technion -- Israel Institute of Technology. D.A. acknowledges the hospitality of the ITP (CAS) and Beihang University, Beijing, China, and a CAS President International Fellowship Initiative (PIFI).

 \vspace{12pt}
\noindent \textbf{Conflict of Interest}

  \vspace{6pt}
\noindent The authors declare no conflict of interest.
  \vspace{12pt}

\noindent \textbf{Keywords}

  \vspace{6pt}
\noindent block copolymer, solvent vapor annealing, defect removal, two-fluid model, Onsager's variational principle

\newpage

%
%

\end{document}